\documentclass[noshowpacs,prl,aps,superscriptaddress,floatfix,twocolumn,footinbib]{revtex4-2} 
\usepackage[T1]{fontenc}

\usepackage[utf8]{inputenc}
\usepackage{lineno,mathtools,url}
\usepackage{soul}
\usepackage{amssymb,mathrsfs}
\usepackage{physics}
\usepackage{graphicx}
\usepackage{hyperref}

\hypersetup{
     colorlinks=true,
     linkcolor=blue,
     filecolor=blue,
     citecolor = orange,      
   }
\usepackage[dvipsnames]{xcolor}
\usepackage{dsfont}
\usepackage{mathbbol} 
\usepackage[normalem]{ulem}
\allowdisplaybreaks
\begin{document}

\title{Heisenberg Uncertainty Inequality and Breaking of Isospin Symmetry in Atomic  Nuclei
}

\author{Sandro Stringari} 

\affiliation{Pitaevskii BEC Center, CNR-INO and Dipartimento di Fisica, Universit\`a di Trento, Via Sommarive 14, 38123 Povo, Trento, Italy}
\affiliation{Trento Institute for Fundamental Physics and Applications, INFN, 38123 Povo, Italy}
\date{\today}

\begin{abstract}
The Heisenberg uncertainty  inequality is used to derive a rigorous lower bound to the amount of isospin impurities in $N=Z$ atomic nuclei, caused by the violation of isospin symmetry. The bound is fixed by the difference between the neutron and proton radii and the sum of the charge exchange monopole strengths. It can be used to check the consistency of advanced many-body calculations accounting for the breaking of isospin symmetry.  The uncertainty inequality is also employed to derive an upper bound to the isovector  dipole moment in terms of  the amount of  isospin impurities, providing an insightful connection between the violation  of parity  and isospin symmetries.

\end{abstract}

\maketitle

After the discovery of the neutron,  Heisenberg   introduced in 1932 \cite{Heisenberg1932} (for the English translation see \cite{Brink1965}) the formalism of isospin to describe neutrons and protons as two distinct quantum states of the same particle, in analogy with the spin formalism. The question whether isospin represents a symmetry of the nuclear  interaction  soon became  of fundamental  interest in both nuclear and particle physics. 
While charge conservation is ensured in isolated systems, reflecting the commutativity of the Hamiltonian with the third  component $T_z$ of the isospin operator ($[H,T_z]=0$), isospin symmetry is violated because of the non commutativity of $H$ with the transverse isospin components ($[H,T_\pm]\neq 0]$) where $T_\pm= T_x\pm iT_y$. The non commutativity implies that the eigenstates of the Hamiltonian are not eigenstates of the isospin operator ${\bf T}^2=T_x^2+T_y^2+T_z^2$, with the consequent occurrence of isospin impurities in the ground state of atomic nuclei. The best known origin of such an effect is  provided by the Coulomb force \cite{BohrMottelson1969}, but the role of   the strong force in breaking isospin symmetry is not yet well understood. Actually, the origin of isospin violation and its consequence on measurable observables is still the object of intense   research activity (for recent discussions and reviews see, for example,
\cite{Satula2018,  Loc2019, ColoMirrors2022, RnRpColo2023, Smirnova23,Sheikh2024Article}  and references therein).    Major motivations are given by the relevance of isospin mixing in the study of 
parity violation effects 
\cite{Donnelly2008}, the availability of novel theoretical and experimental studies concerning charge exchange reactions and the excitation  of the corresponding nuclear giant resonances (see, for example \cite{yako2006,ColoSagawa2014,  Frekers2018,Zegers2023}), the growing interest in  neutron skin effects \cite{yako2006,ColoSagawa2014,Kumar2024}, in isobaric analogue states  \cite{Bentley2007}
and  in general in nuclei far from  stability \cite{AuerbachPR1983,Doba2000,Lee2005}, including Halo nuclei \cite{FREDERICO2012,Signorini2020}, as well as in the $\beta$ decay properties of nuclei close to the $N=Z$ drip line \cite{Kaneko2017,Petrovici2018,Choudhary2025,Algora2025}. Such efforts  also reflect the need  of a better understanding  of the consequences of isospin violation on  the equation of state  of highly asymmetric nuclear matter, including the astrophysical implications for neutron stars (see, for example, \cite{RocaMaza2018,neutronstars24}  and references therein). Recent experiments with high energy collisions of atomic nuclei and in kaon production have in particular pointed out the occurrence of unexpectedly large isospin symmetry breaking effects 
\cite{NA61SHINE:2023azp,Kaon2024}. So it is interesting to identify model independent relations involving the amount of isospin impurities and other physical (and in principle measurable)  quantities.

The main purpose of this work is to provide a rigorous inequality for the isospin fluctuations of quantum origin. We show that  in  $N=Z$ nuclei the inequality provides an explicit lower bound to the amount of isospin impurities present in the ground state  in terms of the difference between the neutron and proton square radii  and of the sum of the charge exchange monopole strengths. We also derive an inequality for the isospin fluctuations in terms of the permanent value of the isovector dipole moment, thereby providing a connection between the violation of parity and isospin symmetries.

Our starting point is the generalized expression  \cite{LevSandro1991,LevSandro1993} \footnote{Equation (\ref{PS}) can be derived (see \cite{LevSandro1993,BECBook2016})  by applying the  Schwartz inequality to the scalar product  between the
operators $A$ and $B$ defined by $(A,B)\equiv\langle A^{\dagger}B\rangle$. This yields the inequality 
 $\sqrt{\langle A^\dagger A\rangle \langle B^\dagger B\rangle} + \sqrt{\langle AA^\dagger \rangle \langle BB^\dagger\rangle} \ge |\langle A^{\dagger}B\rangle|+|\langle BA^\dagger\rangle| \ge |\langle [A^{\dagger},B]\rangle| $. 
Taking the square of this inequality  and using  the property $a+b\ge 2\sqrt{ab}$, with $a=\langle A^\dagger A\rangle\langle BB^\dagger\rangle$ and $b=\langle AA^\dagger \rangle\langle B^\dagger B\rangle$ real and positive numbers,  one finally obtains the main result (\ref{PS}).} 
\begin{equation}
\langle \{A^\dagger,A\}\rangle
\langle \{B^\dagger,B\}\rangle\ge |\langle [A^\dagger, B]\rangle|^2
\label{PS}
\end{equation} 
of the Heisenberg uncertainty inequality \cite{Heisenberg1927,Robertson1929,Schrodinger1930}, providing  rigorous constraints on the  quantum fluctuations of non Hermitian   $A$ and $B$   operators, herafter assumed to have  vanishing average values ($\langle A \rangle = \langle B \rangle=0$). 
Here $\{A^\dagger,A\}\equiv A^\dagger A+AA^\dagger$ and $[A^\dagger,B]\equiv A^\dagger B-BA^\dagger$
are  the usual anticommutator and commutator combinations, respectively.  While the Heisenberg inequality is usually formulated in terms of Hermitian operators (observables), its formulation in terms of non Hermitian operators has been less systematically considered in the literature. In this respect it is worth noticing that, even if the operators $A$ and $B$   are not Hermitian, the corresponding fluctuations represent observables, in many cases of high physical interest.
The derivation of uncertainty inequalities for non Hermitian operators  has actually been the object of several papers in recent years,  stimulated by an increasing interest in non-Hermitian Quantum Mechanics (see for example \cite{Bagarello2023} and references therein \footnote{Differently from Eq.(\ref{PS}),  the uncertainty inequalities considered in these works however contain, in the right hand side of the inequality,  terms depending on the anticommutator   $\{A^\dagger,B\}$ 
rather than on the  commutator $[A^\dagger,B]$ between the relevant operators $A^\dagger$ and $B$}).  

The averages entering Eq.(\ref{PS}) can be taken on  a generic quantum state, usually (but not necessarily) chosen to be the  ground state of the system and the validity of the inequality does not require any specific assumption for the Hamiltonian of the system. In the case of hermitian operators ($A^\dagger=A$ and $B^\dagger =B$) Equation  (\ref{PS}) reduces to the usual uncertainty relation $\Delta A\Delta B\ge (1/2)|\langle [A,B]|$, with $\Delta A\equiv \sqrt{\langle A^2\rangle}$ (and analogously for $B$), first identified by Heisenberg in 1927 \cite{Heisenberg1927} for the position and momentum operators  (see also \cite{Robertson1929,Schrodinger1930}). The inequality (\ref{PS}) was employed in \cite{LevSandro1991,LevSandro1993} to reveal the consequences  of quantum fluctuations in systems characterized by the spontaneous breaking of continuous symmetries and to rule out the occurrence of diagonal long range order in a class of low dimensional systems (Bose-Einstein condensates, antiferromagnets, quantum crystals and fractional quantum Hall effect), thereby providing the zero temperature generalization of the Hohenberg-Mermin-Wagner theorem \cite {Hohenberg1967, MerminWagner1966,Mermin1968},  holding at finite temperature \footnote{The Hohenberg-Mermin-Wagner theorem was actually derived using the Bogoliubov inequality (see \cite{Wagner66}) $\langle\{A^\dagger,A\}\rangle\langle[B^\dagger,[H,B]]\rangle\ge k_B\it{T}|\langle [A^\dagger,B]\rangle|^2$, where $T$ is the temperature, which accounts for the effects of the thermal fluctuations and becomes useless at zero temperature}. In \cite{LevSandro1991,LevSandro1993} the operators $A$ and $B$ were chosen to ensure that the average of the commutator $[A^\dagger, B]$ reflects the explicit signature of the breaking of the corresponding symmetries. In the present work, applied to nuclear systems,  we will employ a similar procedure,  where the isospin symmetry is not broken spontaneously, but by the presence of symmetry  breaking  terms in the Hamiltonian. 

We make the choice $A= T_- \equiv \sum_jt_{-,j}$ and $B= M_-\equiv \sum_jr^2_jt_{-,j}$ for the relevant operators entering the uncertainty inequality (\ref{PS}), where the suffix $j$ runs over the constituents (nucleons) of the nucleus ($j=1,..,A$) and $t_- =t_x-it_y$ is the lowering component of the single particle  isospin operator, transforming a neutron ($t_z=+1/2$) into a proton ($t_z=-1/2)$. Here and in the following the spatial coordinate ${\bf r}_j$ of the $j$-th nucleon refers to the center of mass of the system. 

The choice for the operator $A$ is motivated by the fact that the fluctuation $\{A^\dagger,A\}$ fixes  the total amount 
of isospin for a given value of $T_z$. In fact
$\{T_+,T_-\} = 2({\bf T}^2-T_z^2)$. 
The transverse isovector "monopole" choice for the auxilary operator $B$ is instead  suggested by the form of the mean field  isovector component of
the Coulomb force, which is expected to be a major  responsible for the presence of isospin impurities \cite{BohrMottelson1969}. The fluctuations of the auxiliary operator $B$, fixed by  the  2-body isospin correlation functions,  can be usefully  identified as the sum of the charge exchange isovector monopole strengths relative to the $T_z+1$ and $T_z-1$ channels. In fact, inserting the completeness relation $\sum_n|n\rangle \langle n| = {\bf 1}$ \footnote{The sum should include also the states in the continuum} , one can write $
   \langle \{B^\dagger, B \}\rangle = \sigma^M_+ + \sigma^M_-$,
where   $\sigma^M_+ \equiv \langle M_-M_+ \rangle=  \sum_n|\langle n| M_+ |0\rangle|^2$ and $ \sigma^M_-\equiv \langle M_+M_-  \rangle = \sum_n |\langle n| M_- |0\rangle|^2$ are the static structure factors relative to the monopole charge exchange operators $ M_+ =\sum_jr^2_jt_{+,j}$ and $ M_- = \sum_jr^2_jt_{-,j}$, exciting states in $T_z+1$ and $T_z-1$ channels, respectively.
Notice that the elastic terms $\langle 0 |M_\pm |0 \rangle$  identically vanish if the state where we take the averages,   is   eigen-state of $T_z$. In $N\neq Z$ nuclei the fluctuations of the charge exchange monopole operators are  sensitive to the excitation of the ground state of the isobaric configuration (isobaric analogue state, see also discussion below) \cite{LippariniStringari89,Danielewicz2013}.  

Finally, thanks to the commutation relation $[t_{+,j},t_{-,k}]= 2\delta_{j,k}t_{z,j}$, the  average of the commutator  entering the right hand side of the uncertainty relation  (\ref{PS})  reduces to  $\langle [A^\dagger, B]\rangle
= 2\langle \sum_jr^2_jt_{z,j}\rangle = N\langle r^2_n\rangle -Z\langle r ^2_p\rangle$,
 where  $\langle r^2_n\rangle =(1/N) \int d{\bf r}\rho_n$ and $\langle r^2_p\rangle= (1/Z) \int d{\bf r}\rho_p$
are, respectively, the average neutron and proton square radii. In $N=Z$ nuclei the non vanishing of  this quantity is a direct signature of the breaking of isospin symmetry. Here and in the following the spatial coordinate $\bf{r}$ is always defined in the frame of the center of mass. 

The uncertainty inequality, when applied to the isospin dependent operators introduced above,  eventually takes the  form
\begin{equation}
\langle {\bf T}^2 -T_z^2\rangle \ge \frac{(N\langle r^2_n\rangle-Z\langle r^2_p\rangle)^2}{2(\sigma^M_++\sigma^M_-)}  
\label{S1}
\end{equation}
and represents a non trivial example of rigorous relations predicted by quantum mechanics, involving quantities of relevant physical  interest and holding in complex systems (atomic nuclei) of finite size, where even the exact form of the underlying Hamiltonian is  not known \footnote{Equation (\ref{S1}) can be  equivalently derived employing the Hermitian choice $A= T_x \equiv \sum_jt_{x,j}$ and $B= M_y\equiv \sum_jr^2_jt_{y,j}$ for the operators $A$ and $B$ in the usual formulation of the Heisenberg inequality for Hermitian operators and   using the isospin relationships $t_x=(t_++it_-)/2$, $t_y=(t_+-it_-)/2i$, as well as the explicit assumption that the state $|\rangle$, where we take the averages, is an eigenstate of $T_z$.}. It is also interesting to notice that in spherical nuclei one can use the model independent  sum rule result \cite{Auerbach1982,LippariniStringari89}
$N\langle r^2_n\rangle -Z\langle r^2_p\rangle=3(\sigma^D_+-\sigma^D_-)$ where we have introduced the strengths $\sigma^D_\pm= \sum_n |\langle n| D_\pm|0\rangle|^2$ relative to the   charge exchange   dipole operators $D_\pm =\sum_j z_jt_{\pm,j}$. Using the above relationships one can then express the rhs of the inequality (\ref{S1}) uniquely in terms of  measurable charge exchange strengths of dipole and monopole nature. 

In the  ground state of  $T_z=0$ ($N=Z$) nuclei  the left hand side  of 
(\ref{S1}) would identically  vanish in the absence of interactions violating isospin symmetry, since in this case the ground state  wave function is eigenstate of both $T_z$ and ${\bf T}^2$ with vanishing eigenvalues, ground states  with $T_z=0$ and $T>0$ being ruled out by energetic considerations.  This means that in such nuclei the right hand side of  (\ref{S1}) provides a rigorous lower bound to the amount of isospin impurities caused by the breaking of isospin invariance.  In particular, writing the ground state of such nuclei in the form 
\begin{equation}
|T_z=0\rangle = \sum_{T\ge 0}\epsilon_T|T_z=0, T\rangle \; ,
\label{expansionepsilon}
\end{equation} 
with the normalization condition $\sum_{T\ge 0} \epsilon_T^2=1$, one can write $\langle {\bf T}^2\rangle=\sum_{T\ge 1}\epsilon_T^2T(T+1)$.  By ignoring the negligible isospin symmetry breaking effects in the denominator of (\ref{S1}), and hence setting $\sigma^M_++\sigma^M_-=4\sigma^M_z$, with $\sigma^M_z= \langle M_z^2\rangle = \sum_n\langle |\langle n|M_z|0\rangle|^2$, the static structure factor relative to the isovector monopole operator $M_z=\sum_jr^2_jt_{z,j}$, the uncertainty inequality  eventually takes the simplified form.
\begin{equation}
\langle{\bf T}^2\rangle \ge  
    \frac{N^2(\langle r^2_n\rangle -\langle r^2_p\rangle)^2}{8\sigma^M_z} \; .
\label{S3}
\end{equation}

The inequality (\ref{S3}) can be used to test the consistency of {\it ab initio}
 many-body calculations, accounting for surface effects, full inclusion of the Coulomb interaction and  other interactions violating isospin symmetry \cite{Donnelly2008,Viviani2005,Loc2019}.  In particular, recent investigations of superallowed $\beta$ decay in nuclei close to the $N=Z$ drip line \cite{Kaneko2017,Petrovici2018,Choudhary2025,Algora2025} have pointed out the relevance of proton-neutron pairing correlations which are responsible for a significant contribution to the breaking of isospin symmetry.  It can be also used to test the  consistency of mean field approaches, where the isospin symmetry is additionally spontaneously broken by the use of mean field wave functions \cite{Sheikh2021}. 
It is useful to confirm the validity of result (\ref{S3}) for $N=Z$ nuclei in a simplified model, where one assumes that the violation of isospin symmetry is uniquely provided by the Coulomb interaction for which one retains only the one-body mean field isovector component  $V_{1C}=  \alpha \sum_j  r^2_jt_{z,j}$ , with $\alpha=Ze^2/2R^3$ and  $R$ the radius of a uniform proton density profile \cite{BohrMottelson1969, Loc2019}. Using standard perturbation theory and assuming that the isovector monopole operator $\sum_j  r^2_jt_{z,j}$ just excites  a single state with $T=1$ and  excitation energy equal to $E_M-E_0$,  one finds the following results for 
the quantities entering the  inequality (\ref{S3}) : 
$\langle{\bf T}^2\rangle= 2 \alpha^2 \sigma^M_z/(E_M-E_0)^2$, $N(\langle r^2_n\rangle -\langle r^2_p\rangle)= 2\langle \sum_jr^2_jt_{z,j}\rangle=4 \alpha \sigma^M_z/(E_M-E_0)$. Comparison with Eq.(\ref{S3}) then shows that in this simplified description the inequality (\ref{S3}) reduces to an identity.


Differently from the $N=Z$ case, in $N\neq Z$ nuclei  the transverse isospin fluctuations $\langle \{T_+,T_-\}\rangle$ do not vanish even neglecting the presence  of isospin impurities. So in this case the inequality  (\ref{S1}) is much less sensitive to the amount of isospin mixing. 

Neglecting isospin  impurity effects in the evaluation of $\bf{T}^2$, the   left hand side of Eq.(\ref{S1}),  simply reduces to $\langle{\bf T}^2-T_z^2\rangle = (N-Z)/2$, where we have considered $T_z=T$ nuclei (the formalism is immediately generalized to the case of mirror nuclei with $T_z=-T$). In this case the uncertainty inequality can  be conveniently rewritten in the form
\begin{equation}
N\langle r^2\rangle_n-Z\langle r^2\rangle_p \le \sqrt{(N-Z)(\sigma^M_++\sigma^M_-)} \; .
\label{neq0} 
\end{equation}  
The physics associated with this inequality is connected with the excitation of the isobaric analogue state. Actually neglecting the interaction terms responsible for the violation of isospin symmetry  and noticing that in this case  the isobaric analogue state is exactly given by $|IAS\rangle = (1/\sqrt{N-Z})T^-|0\rangle$,  one finds that the corresponding contribution  $\sigma^M_-({IAS})\equiv|\langle IAS| \sum_j r^2_jt_{j,-} |0\rangle|^2$ to the charge exchange monopole strength $ \sigma^M_-=  \sum_n |\langle n| \sum_j r^2_jt_{j,-} |0\rangle|^2$ takes the form  $\sigma^M_-({IAS})=(N\langle r^2\rangle_n-Z\langle r^2\rangle_p)^2/(N-Z) $ \cite{LippariniStringari89}. In this simplified case  Eq.(\ref{neq0}) takes the trivial form $\sigma^M_{IAS}\le \sigma^M_++\sigma^M_-$.

As already pointed out,  the isovector "monopole" choice for the auxiliary operator $B$ emphasizes the crucial role played by the Coulomb interaction.  Additional bounds for the violation of isospin symmetry can be obtained using different choices for the operator $B$. For example,  adopting the isovector dipole  choice $B= \sum_jz_jt_{-,j}$ rather than the isovector monopole one, one finds a different inequality for the isospin fluctuations involving  the value of the isovector dipole moment calculated on the ground state of the nucleus.  In  this case the Heisenberg uncertainty inequality (\ref{PS}) takes the form  
\begin{equation}
2\langle {\bf T}^2 -T_z^2\rangle(\sigma^D_++\sigma^D_-)\ge d_z^2 \; ,
\label{S4}
\end{equation}
where 
$ \sigma^D_\pm=  \sum_n \langle n|D_\pm |0\rangle|^2$ are the strengths
relative to the charge-exchange dipolar operators $D_\pm=\sum_jz_jt_{j,\pm} $, while $d_z=2\langle \sum_j z_jt_{z,j}\rangle$ is the isovector dipole moment. 

For $N=Z$ nuclei, where  the  isovector dipole moment can be written as  $d_z  =N(\langle z_n\rangle -\langle z_p\rangle)$, the sum of the charge exchange dipolar strengths is accurately approximated by the expression $\sigma^D_++\sigma^D_-=4\sigma^D_z$, with $\sigma^D_z= \langle D_z^2\rangle = \sum_n|\langle n| D_z|0\rangle|^2$  the strength relative to the dipole operator
$D_z=\sum_jz_jt_{z,j}$. The Heisenberg inequality can then be rewritten in the form of the upper bound 
\begin{equation}
d_z^2 \le 8 \langle {\bf T}^2\rangle \sigma_z^D
\label{S5}
\end{equation} 
for the square of  the electric dipole moment $d_z$.
The dipole moment $d_z$ determines the  polarization contribution to the nuclear electric dipole moment, a quantity of high relevance, reflecting the occurrence of violation  of both parity (P) and time reversal (T) symmetry \cite{Pospelov2005,LiuEDMDeuteron2004,vanKolck2011,Yamanaka2017,Navratil2021}.  Eq.(\ref{S5}) then provides an insightful connection between the violation of parity  and isospin symmetries  in $N=Z$ nuclei.  
In particular it proves that  a finite value of the isovector dipole moment implies the occurrence of violation of isospin symmetry. 

In conclusion  we have used the Heisenberg uncertainty inequality to derive a rigorous lower bound to the amount of  isospin impurities in $N=Z$ nuclei, in terms of the difference between the neutron and proton radii, resulting from the breaking of  isospin symmetry. The bound has been obtained starting from  the Heisenberg uncertainty inequality  with a proper choice of non commuting operators.  We have also derived a model independent inequality emphasizing a connection between the breaking of isospin and  parity symmetries.

Further studies will concern a  systematic investigation of the isospin impurities in nuclei close to the $N=Z$ drip line, where the inclusion of important  neutron-proton pairing correlations 
 is required to account for the recently observed superallowed $\beta$ decays \cite{Kaneko2017,Petrovici2018,Choudhary2025,Algora2025}.  The use of optimized choices for the auxilary operator $B$,
accounting  for a more quantitative description of surface and short range effects,  and the role of isospin thermal fluctuations also remain open issues for future studies.
\paragraph{Acknowledgement}
I am grateful to Gianluca Colò for many insightful comments. Stimulating discussions with Winfried Leidemann, Giuseppina Orlandini, Bira van Kolck and Fabio Bagarello are also acknowledged. Old collabotations on the Heisenberg inequality with the late friend Lev Pitaevskii as well as old collabotations on the isospin problem with Enrico Lipparini and the late friend Renzo Leonardi are acknowledged.  This work has been supported by the Provincia Autonoma di Trento. 

\bibliography{Bib}

\end{document}